\documentclass[aps,twocolumn,amsmath,amssymb,showkeys]{revtex4}
\usepackage{times}
\usepackage{graphicx}
\usepackage{dcolumn}
\usepackage{bm}

\begin{document}

\preprint{12th conference on II-VI compounds}

\title{Majority-Carrier Mobilities in Undoped and \textit{n}-type Doped ZnO Epitaxial Layers}

\author{T. Makino}
 \altaffiliation[Present address: ]{Department of Material Science, University of Hyogo, Kohto, Kamogoori-cho 678-1297, Japan}
 \email{makino@sci.u-hyogo.ac.jp}
\author{Y. Segawa}%
\affiliation{%
Photodynamics Research Center, Institute of Physical and Chemical Research (RIKEN), Sendai 980-0845, Japan}%

\author{A. Tsukazaki}
\author{A. Ohtomo}
\author{M. Kawasaki}
\affiliation{%
Institute for Materials Research, Tohoku University, 2-1-1 Katahira, Aoba, Sendai, 980-8577, Japan}%

\date{\today}

\begin{abstract}
Transparent and conductive ZnO:Ga thin films are prepared by laser molecular-beam epitaxy. Their electron properties were investigated by the temperature-dependent Hall-effect technique.
The 300-K carrier concentration and mobility were about $n_s \sim 10^{16}$~cm$^{-3}$ and 440~cm$^{2}$/Vs, respectively.
In the experimental `mobility vs concentration' curve, unusual phenomenon was observed, i.e., mobilities at $n_s \sim 5\times$ 10$^{18}$ cm$^{-3}$ are significantly smaller than those at higher densities above $\sim 10^{20}$ cm$^{-3}$. Several types of scattering centers including ionized donors and oxygen traps are considered to account for the observed dependence of the Hall mobility on carrier concentration. The scattering mechanism is explained in terms of inter-grain potential barriers and charged impurities. A comparison between theoretical results and experimental data is made.
\end{abstract}

\maketitle

\section{Introduction}

Semiconducting materials with a wide band gap, such as SnO$_2$ and In$_2$O$_3$, are commonly used as transparent electrodes in optoelectronics and solar energy conversion technology. Recently, it has been demonstrated that ZnO:Ga may be considered as the next attractive transparent and conductive oxide compounds.

ZnO has a natural tendency to be grown under fairly high-residual \textit{n}-type in which high concentration of charge carriers of about 10$^{22}$ cm$^{-3}$ may be achieved with group-III element doping. Therefore, highly conductive ZnO films are easily prepared.

To date, there have only been a few attempts~\cite{ellmer_mob1,miyamoto_mob1} to model electron transport in \textit{doped} ZnO. The present publication reports carrier concentration dependence of mobility including an effect of the scattering by ionized donor centers as well as by the grain boundaries~\cite{fischetti_mob1,lowney_mob1}.

\section{Experimentals}

Our samples were grown on lattice-matched ScAlMgO$_{4}$ (SCAM) substrates by laser molecular-beam epitaxy. ZnO single-crystal and (Ga,Zn)O ceramics targets were ablated by excimer laser pulses with an oxygen flow of $1\times 10^{-6}$~torr ~\cite{ohtomo_sst,makino_sst}. The films were patterned into Hall-bars and the contact metal electrodes were made with Au/Ti for \textit{n}-type film, giving good 
ohmic contact.

\section{Results and discussion}

The ZnO materials parameters used in the transport theory calculation has been given elsewhere~\cite{makino_trans}. We have not adopted the relaxation
time approximation for the mechanisms that involve relatively 
high-energy transfers, \textit{e.g.}, longitudinal optical 
phonons. Since Rode's iterative technique takes a long time to reach its convergence~\cite{rode_bkmob1,fischetti_mob1}, the present computations 
are based on the variational principle method~\cite{lowney_mob1,ruda_mob1}. The following electron scattering mechanisms are considered: (1)
polar optical-phonon scattering, (2) ionized-impurity 
scattering, (3) acoustic-phonon scattering through the deformation 
potentials, and (4) piezo-electric interactions~\cite{seeger_bkmob1,lowney_mob1,ruda_mob1}.

The values of $\mu $ are 440~cm$^{2}$/Vs and 5,000~cm$^{2}$/Vs at 300 and 100~K, respectively. We have derived partial mobilities by accounting for respective 
scattering mechanisms in the nondegenerate (undoped) limit. The total electron mobility was calculated by combining all of the partial scattering mechanisms. Our experimental data are in reasonably good 
agreement with theory. The mobility limit at 300~K is about 430~cm$^{2}$/Vs.

On the other hand, the situation is somewhat different for the 
cases of Ga-doped n-type films~\cite{makino_int_Ga}. Figure~1 shows 300-K experimental
mobilities plotted against carrier concentration ($n$).
\begin{figure}[h]
\includegraphics[width=.47\textwidth]{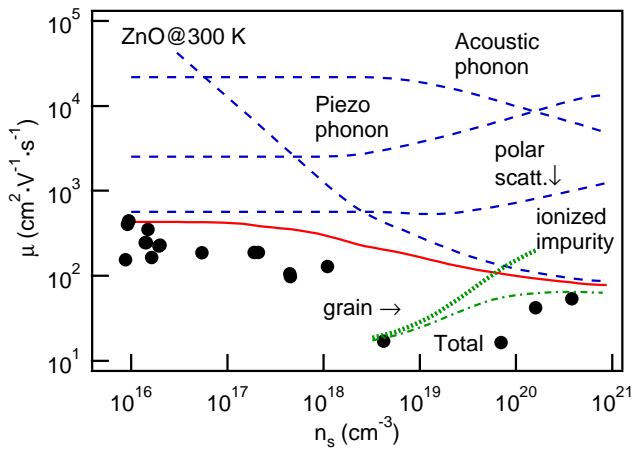}
\caption{Comparison of drift mobility calculations (solid red
curve) with the Hall effect measurements for undoped and doped 
epitaxial films (filled black circles). The contributions of various scattering 
mechanisms to the total mobility are also shown by dashed blue curves.
Also shown are the best fit (dash-dotted green curve) to the experimental data with with contribution from scattering at grain boundaries ($\mu_b$, dotted green curve).} \label{fig:1}
\end{figure}
The mobilities
of doped films are significantly smaller than those of undoped
one~\cite{makino18,makino19}. Obviously, this tendency can be qualitatively attributed to the increased density of impurities. For quantitative comparison,
partial mobilities are calculated and given in Fig.~1 by dashed
lines. We have taken the effects of screening for both ionized impurities and polar optical phonon scattering into account. Polar interactions reflecting the ionicity of
the lattice are dominant in scattering mechanism, while, at heavier doping
levels, ionized impurity scattering controls inherent mobility
limit curve~\cite{ruda_mob1,lowney_mob1}. The experimental data agree well with our calculation (solid curve)
except for the intermediate concentration range. Particularly, our model could not reproduce
a relative minimum in the ``$\mu $ vs $n_s$'' curve, which
has been experimentally observed in this intermediate doping
range. The situation
at $n_s$ \texttt{>} 10$^{20}$ cm$^{-3}$ could be improved probably if the 
effects of non-parabolic band structure as well as of clustering 
of charged carriers would be taken into account~\cite{ellmer_mob1}.

The presence of grain boundaries and trapped interface charges in semiconductors leads to inter-grain band bending and potential barriers~\cite{pisarkiewicz1}. Under specific conditions, this effect may be so prominent that it can significantly influence the scattering process of free carriers, giving rise to a considerable reduction in the Hall mobility.

It is well established that grain boundaries contain fairly high density of interface states which trap free carriers from the bulk of the grains. Such a grain may be thought of as a potential barrier for electrons characterized by its height $E_b$ and width $\delta$ for a given number $Q_t$ of traps per unit area.

The contribution $\mu_b$ (boundary partial mobility) to the total mobility $\mu_T$ that comes from the scattering at grain boundaries is thermally activated and can be described by the well-known relation:

\begin{equation}
\mu_b = \mu_0 \exp (-\frac{E_b}{kT}).
\end{equation}

With $mu_0 = L_q (8\pi m^* kT)^{(-1/2)}$, where $q$ is the charge of a trap (in this case $q=e$). By solving the Poisson equation, one obtains:

\begin{equation}
\label{eq:barrierheight}
E_b = \frac{Q_t}{8 \epsilon_0 \epsilon_s N_d},
\end{equation}
where $N_d$ is the concentration of ionized donors and other nomenclatures take
conventional meanings.

Therefore in our model, we have two free parameters, i.e., $Q_t$ and $\delta$, that can be determined from the best fit of

\begin{equation}
\mu_T^{-1} = \mu_b^{-1}+ \mu_{nb}^{-1},
\end{equation}
to the experimental data $\mu_H(n)$, where $\mu_{nb}$ (the full curve in Fig. 1) refers to all the partial mobilities except for $\mu_b$.

We calculated the barrier height $E_b$ according to Eq.~(\ref{eq:barrierheight}). At low electron concentrations $E_b$ is large, so we can infer that in that case the Hall mobility is barrier limited.

We compared our theoretical results with those experimentally determined. The best fit of our model including all of the contributions to the experimental dependence of $\mu_H$ on n for ZnO films is presented in Fig.~1. The best fit yields $7.5\times 10^{13}$ $cm^{-2}$ and $\delta =$ 2~nm.

Partial contribution from $\mu_b$ is separated by a dash-dotted curve in Fig.~1 to give a better understanding of the scattering mechanism. The contribution from barrier-limited scattering is of no importance as far as high electron concentrations are concerned. Unfortunately, our model can be applied only to the case of degenerate semiconductor regime.

\section{Summary}

The electrical properties of wide band gap, degenerate ZnO thin films are investigated. The experimental dependences of the Hall mobility on the electron concentration is explained mainly in terms of scattering by grain boundaries and charged impurities. Oxygen accumulated in the grain boundaries plays an important role in the scattering mechanism.

\section*{acknowledgements}
  One of the authors (T. M.) thanks H. S. Bennett of NIST, D. L.
Rode of Washington University, St. Louis, USA, H. Ruda of University
of Toronto, Canada, and B. Sanborn of Arizona State University, USA for helpful
discussion. Thanks are also due to Shin Yoshida for technical
assistance during our experiments.


\end{document}